\documentclass[10pt]{article}

\usepackage{amsmath}
\usepackage{amssymb}

\usepackage{graphicx}

\usepackage{cite}

\usepackage{color} 

\usepackage[labelfont=bf,labelsep=period,justification=raggedright]{caption}

\makeatletter
\renewcommand{\@biblabel}[1]{\quad#1.}
\makeatother

\date{}

\begin{document}

\baselineskip 18pt

\begin{flushleft}
{\Large
\textbf{Parametric pattern selection in a reaction-diffusion model}
}
\\
Michael Stich$^{1,\ast}$, 
Gourab Ghoshal$^{1}$, 
Juan P\'erez-Mercader$^{1,2}$
\\
\bf{1} Department of Earth and Planetary Sciences,
Harvard University,
100 Edwin H. Land Boulevard,
Cambridge, MA  02142-1204, USA.\\
\bf{2}  The Santa Fe Institute, 1399 Hyde Park Road, Santa Fe, NM 87501, USA.
\\
$\ast$ E-mail: mstich@fas.harvard.edu
\end{flushleft}

\section*{Abstract}
We compare spot patterns generated by Turing mechanisms with those
generated by replication cascades, in a model one-dimensional
reaction-diffusion system.  We determine the stability region of spot
solutions in parameter space as a function of a natural control
parameter (feed-rate) where degenerate patterns with different numbers
of spots coexist for a fixed feed-rate.  While it is possible to
generate identical patterns via both mechanisms, we show that
replication cascades lead to a wider choice of pattern profiles that
can be selected through a tuning of the feed-rate, exploiting
hysteresis and directionality effects of the different pattern
pathways.


\section*{Introduction}

Reaction-diffusion systems are well known to self-organize into a
variety of spatio-temporal patterns including, spots, stripes,
spirals, as well as spatio-temporal chaos and uniform
oscillations~\cite{Cross09,MikSynI94,Walgraef97}.  Their existence in
out-of-equilibrium states, connection to idealized chemical systems,
and dependence on dimensional parameters, make them a good testbench
for the study of general features of pattern generation and evolution.
In particular, the dependence of these final states on the rate at
which constituents are fed into the system (feed-rate) is of
significant interest, since reaction-diffusion systems represent
proxies for high-level biological systems that can exchange matter
and energy with the environment ~\cite{Grzybowski09}. Depending on the
value of the feed-rate, the system may asymptote into one of many
states and thus the feed-rate can be thought of playing the role of a
natural control parameter.

While spatio-temporal patterns in reaction-diffusion systems (like
replicating spots~\cite{LeeN94} and Turing
patterns~\cite{CastetsPRL90,OuyangN91}) have been found and discussed
extensively in the context of chemical
systems~\cite{Walgraef97,DeWitACP99}, their phenomenology is
ubiquitous.  A well-studied example from physics is related to
electrical current filament patterns in planar gas-discharge
systems~\cite{AstrovPRL97,AstrovPLA06}. The system dynamics can be
described by activator-inhibitor reaction-diffusion models and
different mechanisms of spot array formation have been observed:
division and self-completion.  The relevant control parameter in this
system is the feeding voltage.  Another example that have attracted
interest recently is found in the realm of fluid dynamics where
``spots'' of turbulent regions in pipe flow~\cite{AvilaS11} and plane
Couette flow~\cite{ShiPRL13} have been observed: On a laminar
background, patches of localized turbulence, called puffs, emerge via
finite-amplitude perturbations and also show splitting behavior. These
systems have been recently mapped onto excitable reaction-diffusion
systems~\cite{BarkleyPRE11}, and subsequently, the Turing mechanism
has been proposed to explain the periodic arrangement of puffs
in~\cite{MannevilleEpL12}, suggesting again a reaction-diffusion
framework for the dynamics. The corresponding control parameter in
this case is the Reynolds number of the flow.

\begin{figure}[h]
\centering 
\includegraphics[width=0.5\textwidth]{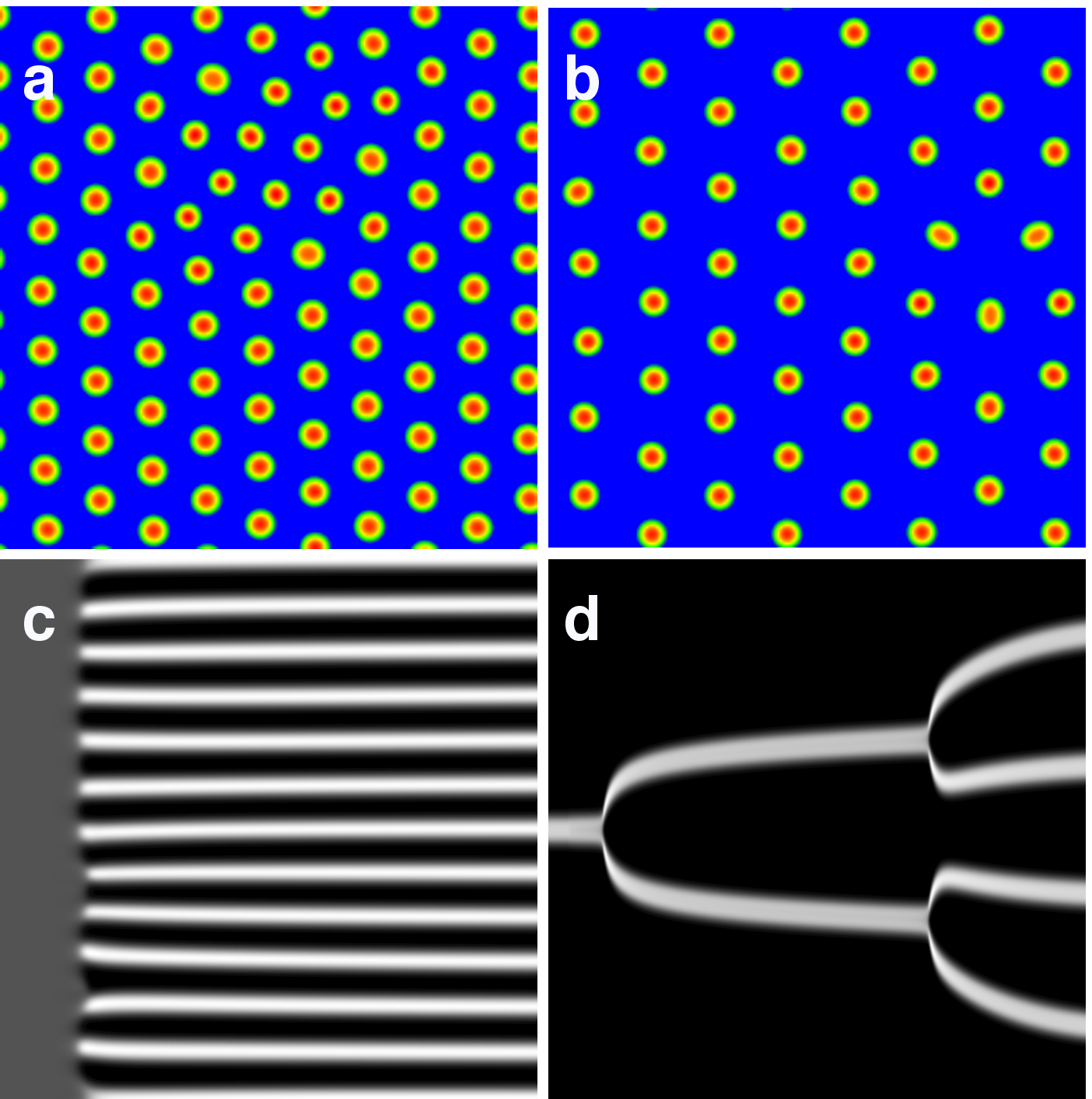}
\caption{{\bf Stable stationary spot arrays in the reaction-diffusion
  system~(1) generated by (a) Turing instability, (b) replication
  cascade.} Two space dimensions are considered, with system size
  $L_x=L_y=100$ and periodic boundary conditions. Typical formation
  pathways for the Turing case (c) and the replication scenario (d)
  are shown in the space-time diagrams for simulations in
  one-dimensional space with $L=150$.  In (a-d), the variable $a$ is
  displayed in color code: red, respectively white denote large
  values. Parameters: (a) $F=3.00$; (b) $F=2.20$; (c) $F=3.00$,
  displayed time interval $200$, (d) $F=2.49$, displayed time interval
  $3000$.  Other parameters as in Fig.~\ref{fig:fig333}. A pattern
  profile for both variables $a$ and $b$ will be shown in
  Fig.~\ref{fig444}(b).}
\label{fig:fig111}
\end{figure}

While these examples show that similar phenomena appear in
\emph{different systems}, an even more intriguing feature is that
patterns that look qualitatively similar can be generated by very
\emph{different mechanisms} in the \emph{same system}.  Consider the
patterns shown in Figures~\ref{fig:fig111}(a,b), which are the result
of numerical simulations of a typical bistable reaction-diffusion
system in two spatial dimensions.  While both figures represent
stationary arrays of spots (increased concentrations of one or more
chemical species relative to others), their evolutionary pathways are
quite different.  Figure~\ref{fig:fig111}(a) was generated by the
Turing mechanism~\cite{Turing52}, i.e.  from a uniform stationary
state unstable under spatial perturbations, giving rise to a
stationary, spatially periodic pattern. This is illustrated by a
space-time diagram for a simulation in one space dimension in
Fig.~\ref{fig:fig111}(c), where an initially uniform state almost
simultaneously develops $n$ spots as a result of the small random
perturbation.

In contrast to the above, the pattern in Fig.~\ref{fig:fig111}(b) was
generated by perturbing a \emph{different} uniform steady state,
creating a single spot, that after a slight increase in the feed-rate,
undergoes a replication cascade of spots, eventually filling the space
(again illustrated in Fig.~\ref{fig:fig111}(d) by a space-time diagram
for a simulation in one space dimension).  Thus, while the asymptotic
state of the system looks similar in both cases, the initial
conditions, the parameter regimes in which they occur, and the
mechanisms by which they are generated are different.  

In the face of this, it is of interest to investigate if there is an
abrupt transition or a smooth continuation --as a function of the
feed-rate-- between the patterns, as one traverses from one limit to
the other. If there is a coexistence region, we want to investigate
whether the asymptotic states of these patterns are identical and only
the temporal evolution differ.  Finally, since it may be desirable to
select particular states of the system we seek to determine if it is
possible to \emph{use} the different mechanisms to smoothly engineer
transitions between different states.

In this article, we explore these questions in a model
reaction-diffusion system that displays both replication cascades
\emph{and} Turing instabilities.  In the spirit of simplicity,
tractability and clarity, we focus on a medium with only one spatial
dimension and investigate the formation of patterns as a function of
the feed-rate $F$. Therefore, we do not consider other pathways for
the generation of spot solutions, such as transverse instabilities of
stripe solutions (requiring at least two spatial dimensions).  We find
that, while the mechanisms driving the formation of spot arrays are
discernibly separated in different regimes of $F$, the patterns are
essentially indistinguishable in intermediate regimes.  Nevertheless,
we find degeneracies, hysteresis and directionality effects that can
be exploited for the purposes of pattern selection, via the tuning of
the feed-rate.

\section*{Results}

\subsection*{The model and basic instabilities}

Our model reaction-diffusion system (first introduced in~\cite{LesmesPRL03}) is
described by the differential equations
\begin{subequations}
\begin{align}
\frac{\partial a}{\partial t} &= k_1 a^2 b - k_2 a + D_a \frac{\partial^2 a}{\partial x^2}, \\
\frac{\partial b}{\partial t} &= - k_1 a^2 b - k_3 b + F  + D_b \frac{\partial^2 b}{\partial x^2}, 
\end{align}
\label{eq:diffeqmodel2}
\end{subequations}
where $a$ can be interpreted as the concentration of an activator
$\textrm{A}$ and $b$ as the concentration of a substrate $\textrm{B}$.
There is an autocatalytic step for $\textrm{A}$ at rate $k_1$, and
decay reactions for $\textrm{A}$ and $\textrm{B}$ at rates $k_2,k_3$,
while $\textrm{B}$ is fed in to the system at a rate $F$.  The model
is closely related to a class of well-studied reaction-diffusion
systems such as the~Sel'kov-Gray-Scott
model~\cite{SelkovEJBC68,GrayCES83,PearsonS93} (see also Supplementary
Information, Sec. S1), the Gierer-Meinhardt model~\cite{GiererK72} and
the Brusselator~\cite{PrigogineJCP68}.

We begin our analysis by first determining the uniform absorbing
states of the model and then proceed to determine the specific
instability associated with each state.  Without loss of generality,
the concentrations can be rescaled to $a \to \sqrt{k_1}a~\textrm{and}~b
\to \sqrt{k_1} b$; then the stationary uniform states are determined
by setting the right-hand side of Eq.~(\ref{eq:diffeqmodel2}) to zero.
Doing so, we obtain $(a_1,b_1)=(0,F/k_3)$, which we refer to as state
{\bf 1}. At a critical value of the feed-rate $F_{SN}=2k_2 k_3^{1/2}$,
we find that two more solutions are generated by a saddle-node
bifurcation. The first is an \emph{unstable} intermediate state~{\bf
  2} and the second is a \emph{stable} state~{\bf 3} given by
$(a_3,b_3)=((F+\sqrt{F^2-4k_2^2k_3})/2k_2,(F-\sqrt{F^2-4k_2^2k_3})/2k_2)$.
In addition to this we find that the system undergoes a Hopf
bifurcation at yet another critical value $F_{H} = k_2 ^2
/\sqrt{k_2-k_3}$, whereby in the range $F_{SN} \le F \le F_{H}$, state
{\bf 3} is potentially unstable with respect to temporal oscillations
(for details see Supplementary Information, Sec. S2).

Thus, the primary absorbing states of interest are {\bf 1} and {\bf
  3}. These turn out to display distinct forms of instability.  At a
critical feed-rate $F=F_T$, state {\bf 3} is linearly unstable with
respect to spatially inhomogeneous perturbations, leading to the
formation of Turing patterns in the interval $F_{SN} \le F \le F_T$
(see Supplementary Information, Sec. S3).  The characteristic
wavelength of the pattern $\lambda$ can be determined through a
standard linear stability analysis, and this determines the
total number of spots $n$ that are present in the system through the
simple relation $L=n\lambda=2 \pi n/k$, where $L$ is the system size
and $k$ the wavenumber (see Methods).

On the other hand, while state~{\bf 1} is stable with
respect to infinitesimally small perturbations, it is \emph{unstable}
to localized large-amplitude perturbations, that can induce the
formation of a single spot.  Using the technique of scale-separation,
one can calculate the profiles of the spot solutions, along with the
parameter regimes for which they exist.  In a particular limit, where
$(k_3 D_a/ k_2 D_b)^{1/2} \ll 1$, we can define a critical feed-rate
for the formation of single-spot solutions, such that spots exist for
$F \ge F_{sp}=2 \sqrt{3} (k_2 k_3)^{3/4} (D_b/D_a)^{1/4}$ (details are
shown in Supplementary information, Sec. S4).  As $F$ is further
increased, the single spot becomes unstable with respect to a
replication cascade (at a numerically determined critical feed-rate
$F_{rep}$) which eventually fills the system with a spot
array, for related work for the Gray-Scott model see~\cite{ReynoldsPRE97,DoelmanNl97,NishiuraPD99,MorganMAA00,MuratovJPAMG00,EiJJIAM01,MuratovSIAMJAM02,KolokolnikovPD05,WeiJMB08}.

It is essential to point out that the fundamental difference between
the formation of spot arrays via the Turing mechanism or via
replication cascades, is that the former results from an instability
of state~{\bf 3} to infinitesimally small perturbations with a
characteristic wavelength, while the latter is the result of a
localized large-amplitude perturbation to state~{\bf 1}.

\subsection*{Turing patterns and localized spot patterns}
\label{sec:localTuring}

We next investigate the differences between these two pattern
formation mechanisms through the aid of numerical simulations, where
we initialize the system in a variety of different initial states and
examine the corresponding asymptotic states.  To compare the generated
patterns, we need to choose a suitable metric to distinguish them. In
principle, there are many quantities one can measure, however, as
Fig.~\ref{fig:fig111} suggests, a particularly simple choice would be
to simply count the number of spots $n$ that are generated in the
asymptotic state of the evolution of the system.

\begin{figure}[h]
\centering 
\includegraphics[width=0.6\textwidth]{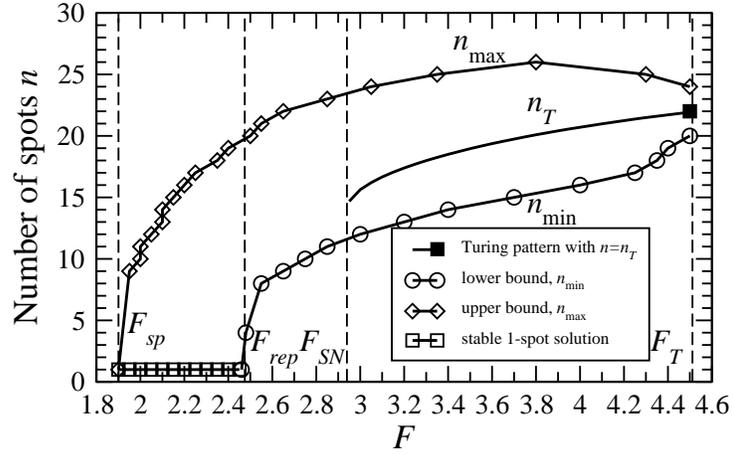}
\caption{{\bf Stability area for $n$-spot arrays as a function of $F$ for a
  system size $L=200$ with periodic boundary conditions and $k_2 =
  1.3, k_3 = 1.5, D_a=1, D_b=50$ (details of simulation are covered in
  Methods).}  The stability area is enclosed by the curves
  $n_{\textrm{max}}$ and $n_{\textrm{min}}$, corresponding to the
  maximum and minimum number of stable spots for a given $F$.  $F$ is
  changed in steps of $\Delta F=\pm 0.05$ using the asymptotic state
  of the previous $F$ as initial condition (ramping).  Turing patterns
  are marked by the curve $n_T$.  Vertical lines correspond to the
  values for the instabilities: $F_{sp}=1.90$, $F_{rep}=2.45$,
  $F_{SN}=2.94$, $F_{T}=4.51$. }
\label{fig:fig333}
\end{figure}

Consider the plot in Fig.~\ref{fig:fig333}, where we show the number
of spots $n$ as a function of the feed-rate $F$ in the asymptotic
state of the simulation (for the numerical details of the simulations,
see Methods).  We start with a single spot induced on the background
of state~{\bf 1} in the region $F \ge F_{sp}$ (that supports stable
spots) and gradually increase $F$ in small increments of $\Delta
F$. Doing so, we eventually reach a critical $F_{rep}$ where the spot
splits into two spots (replicates). The 2-spot solution may again be
unstable, and the splitting process is repeated.  This is the
situation if we start with a \emph{single} spot as initial condition.
However, in the region $F \ge F_{sp}$, we can also directly create a
$n$-spot array with $n > 1$ by inducing multiple large amplitude
perturbations in different spatial locations of the system.  The size
of each spot of course is finite (being determined by the diffusion
coefficients $D_{a,b}$) and consequently there is a maximum number of
spots $n_{\textrm{max}}$ that can be supported in a finite medium.
Thus in the region $F_{sp} \le F \le F_{rep}$ we can initialize a wide
range of spot arrays within the bounds $1 \le n \le n_{\textrm{max}}$
and by the same procedure of incrementing $F$, determine the values of
$F$ at which the spot array replicates.  The resulting curve is
displayed in Fig.~\ref{fig:fig333} as the lower boundary of the
stability area. These values of $F$ for each $n$ represent a
generalization of the critical feed-rate $F_{rep}$ for $n >1$.
Clearly, this also implies that the curve corresponds to the
\emph{minimum} number of stable spots $n_{\textrm{min}}$ that can be
supported by the system for fixed $F > F_{rep}$, and we thus label
this curve $n_{\textrm{min}}(F)$.

We next turn our attention to the Turing regime ($F_{SN} \le F \le
F_{T}$) and the spot patterns found there. The onset of the Turing
instability is of special interest: by inducing a small-amplitude
perturbation around state~{\bf 3} at $F = F_T$, we obtain a
\emph{native} Turing pattern of $n_T=22$ spots (denoted in
Fig.~\ref{fig:fig333} with a black square) in very good agreement with
the theoretical value predicted by linear stability analysis (see
Supplementary information, Sec.~S3 and Eq.~(S9)).  Away from $F_T$,
the analysis provides us with a continuous \emph{band} of unstable
wavelengths.  Extensive simulations show that in the entire Turing
regime ($F_{SN} \le F \le F_{T}$), small random perturbations of
state~{\bf 3} lead \emph{on average} to a spot pattern with $n_T$
spots (marked by the solid curve extending from $n_T=22$ at $F = F_T$
in Fig.~\ref{fig:fig333}), as predicted by linear stability analysis
using the \emph{most unstable} wavelength.  This is in agreement with
similar findings for the Gray-Scott model~\cite{MazinMCS96},
confirming that patterns in this regime and initialized in this way
are indeed bonafide Turing patterns.

Comparing the replication mechanism with the Turing mechanism,
we recognize that the former provides an elegant
way to access a number of spots that are \emph{different} from $n_T$
(the native Turing pattern) \emph{within} the Turing regime\footnote
{We note that Turing patterns with $n \ne n_T$ can also be generated
  by expanding fronts generated by perturbing state~{\bf 1} in the
  Turing regime (see Supplementary Information, Sec.~S5), however this
  is not the focus of this article.}. This is done by first
initializing a $n$-spot pattern for $F_{sp} \le F \le F_{SN}$ (outside
the Turing regime), and then gradually increasing $F$ until we are
within the Turing regime. In this way we can select a wide range of
$n$ within the bounds $n_{\textrm{min}} \le n \le n_{\textrm{max}}$
that differ from $n_T$.

Furthermore, starting from any stable $n$-spot array, we are free to
reverse the procedure and \emph{decrease} $F$ in increments of $\Delta
F$. We find that after a particular value of $F$ is reached, $n$ now
\emph{decreases}.  By continuing this process and repeating it for all
$n$, we obtain the upper curve in Fig.~\ref{fig:fig333} that gives the
\emph{maximum} number of stable spots $n_{\textrm{max}}$ that can be
sustained for a given $F$.  The area enclosed by the curves
$n_{\textrm{min}}$ and $n_{\textrm{max}}$ thus marks the stability
region of $n$-spot arrays as a function of the feed-rate $F$.  We
immediately see from the figure that degenerate $n$-spot arrays exist
for a large range of $F$, where the arrays can in principle be
generated by \emph{different} mechanisms.

Taken together, these results allow us to interpret $n_{\textrm{max}}$
as a \emph{disappearance} boundary where a $n$ spot solution goes to a
new value $n' < n$, and $n_{\textrm{min}}$ as a \emph{splitting}
boundary where $n' > n$. In general, in an infinite system, $n$ spots
split into $2n$ spots, however in a finite system this is constrained
by its size. Therefore even in the region that supports replication,
for large enough $n$, some of the spots in the array splits while
other do not.  The specific value of $n'$ is sensitive to
small perturbations, in particular at the moment the splitting or
disappearance takes place.

\begin{figure}[h]
\centering 
\includegraphics[width=0.6\textwidth]{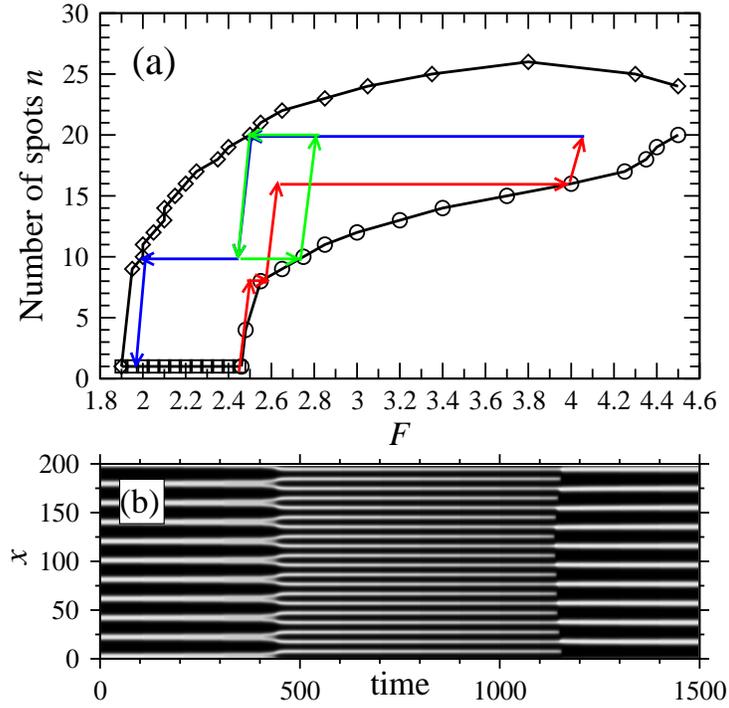}
\caption{{\bf Different pattern pathways.} (a) The red and blue
  pathways represent a hysteresis curve for an example $n$-spot array
  induced in state~{\bf 1}. We observe a sequence $1 \to 8 \to 16 \to
  20 \to 10 \to 1$ spots. The green path represents a cycle between 10
  and 20 spots (more see text). (b) Space-time diagram for $a$ along
  the green path shown in (a). $F$ is changed about $\Delta F=0.05$
  each $\Delta t=100$. Simulation starts with a 10-spot solution at
  $t=0$ with $F=2.60$, and $F$ increases until $F=2.80$, where
  splitting is observed. Then $F$ is decreased until $F=2.45$, where
  10 spots disappear, and after which it is increased again until
  $F=2.60$ is reached.}
\label{fig:fig333b}
\end{figure}

Clearly, as we can create many different initial conditions, 
many different splitting or disappearance 
pathways exist. As an illustrative example we
show one where a single spot is initialized on the background of state
{\bf 1}.  By increasing $F$, the solution reaches the boundary
$n_{\textrm{min}}$ and splitting ocurrs. The resulting two spots are
also unstable, and finally an 8-spot array is formed. By further
increasing $F$, the array splits into 16 spots. Then it maintains
stability for a wide range as $F$ is increased further, well into the
Turing regime, until it splits as it encounters $n_{\textrm{min}}$
again. This evolution is shown via the red path in
Fig.~\ref{fig:fig333b}(a).  If we now decrease $F$, the boundary
$n_{\textrm{max}}$ is encountered twice, and finally the number of
spots decreases to 1 again (shown as a blue path in
Fig.~\ref{fig:fig333b}(a)). This is an example of a hysteresis curve
connected to the degeneracy of the $n$-spot arrays.

Another example is shown by the green path in
Fig.~\ref{fig:fig333b}(a), where we cycle the spot-array solution
between 10 and 20 spots. To illustrate how this cycle lookes in a real
simulation, in Fig.~\ref{fig:fig333b}(b) we show a space-time diagram
for the variable $a$ along the green path. We start from a 10-spot
solution for $F=2.60$, increase $F$ in small steps until splitting to
a 20-spot solution is observed. Then, we decrease $F$ while preserving
the 20-spot solution (hysteresis) until finally disappearance of spots
takes place and the 10-spot solution is recovered.

The hysteresis effect clearly has the consequence of a preferred
directionality in the system for inducing a replication pathway.
Replication cascades proceed only via an \emph{increase} in the
feed-rate and for $F_{rep} \le F \le F_{T}$. Conversely, the formation
of a Turing pattern appears for a \emph{fixed} $F_{SN} \le F \le
F_{T}$ and for a particular class of initial conditions.

\subsection*{Pattern profiles of $n$-spot solutions}

\begin{figure*}[t]
\centering 
\includegraphics[width=\textwidth]{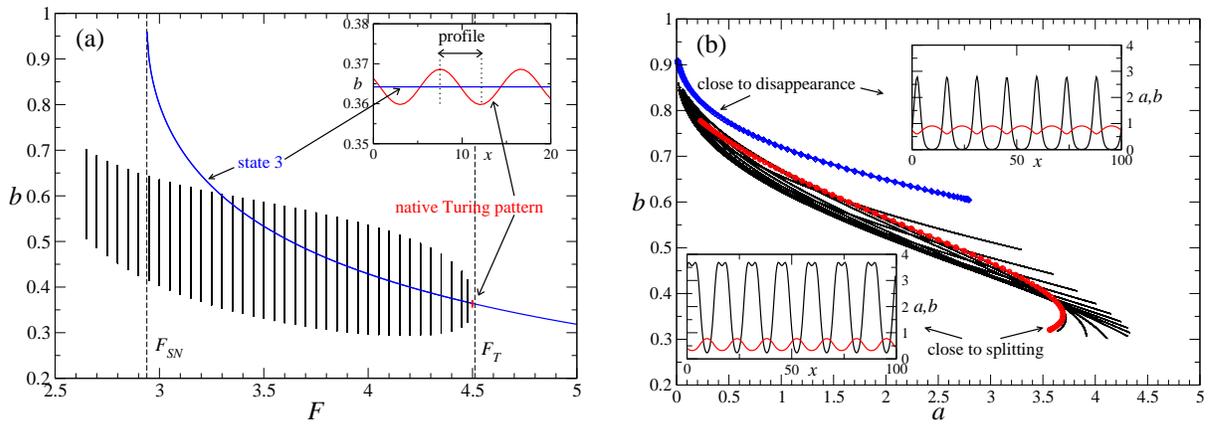}
\caption{{\bf Pattern profiles.} (a) The profile of $b$ in a Turing
  pattern as a function of $F$ for fixed $n=22$. The blue curve
  represents the steady-state value $b_3$. The vertical dashed lines
  $F_{SN}$ and $F_T$ mark the Turing regime.  A Turing pattern appears
  supercritically at $F_T$ (inset) and its amplitude increases as one
  moves away from threshold. At a certain point, the profile ceases to
  oscillate around $b_3$, and continues to exist beyond the Turing
  regime without qualitative changes.  (b) Multiple $14$-spot profiles
  in $(a,b)$ space, for $F$ between $n_{\textrm{max}}$ at $F=2.10$
  (blue curve) and $n_{\textrm{min}}$ at $F=3.42$ (red curve). The
  insets show the corresponding concentration profiles (black is $a$,
  red is $b$).}
\label{fig444}
\end{figure*}

While measuring $n$ has been fruitful in determining the stability
region of the solutions, it does not provide any detailed information
about the spatial distribution of the pattern.  Do the spot arrays
created by Turing instability and spot arrays created by replications
show any differences? Clearly, as one changes $F$ smoothly, the
distribution of the concentration will vary, even as $n$ remains
constant.  A simple way of determining this is to measure the
\emph{profile} of the spots, which is the spatial range between its
maximum and minimum concentrations. A visual illustration of this
definition is provided in Fig.~\ref{fig444}(a) inset.

To investigate this, we initialize a pattern with $n_T=22$ spots at
the Turing boundary $F_T$ and examine the change in profile as we
\emph{decrease} $F$.  In Fig.~\ref{fig444}(a), we plot the profile of
$b$ in function of $F$.  In the same figure we mark the existence
region of state {\bf 3} by the dashed vertical boundary $F_{SN}$ as
well as the steady-state value $b_3$ by a solid blue curve (note that
state~{\bf 3} exists only for $F > F_{SN}$).  We find that close to
$F_T$ the amplitude of the pattern (marked by the vertical solid
lines) is small, and the concentration of $b$ oscillates symmetrically
around state~{\bf 3}, in line with what is expected at $F_T$ (see
inset).  However, as we move away from $F_T$, the amplitude increases
and the profile shifts in phase space.  At some point the pattern
ceases to oscillate around state~{\bf 3}, and eventually decouples
from state~{\bf 3}, continuing to persist even \emph{below}
$F_{SN}$. This decoupling occurs without any qualitative change as the
pattern crosses the boundary.  The implication of this is that for
$F<F_{SN}$ the persistent spot pattern can be interpreted as a
continuation of a Turing pattern, although it is independent of
state~{\bf 3} (unlike a near-threshold Turing pattern) and no Turing
analysis can be applied.  In fact, spot arrays with same $n$, but
created either through the Turing mechanism or replication cascades
show no quantitative or qualitative difference, implying that arrays
created by the two mechanisms are practically indistinguishable in the
intermediate regime.

We next examine the change in profile as we vary $F$ between the
stability boundaries $n_{\textrm{max}}$ and $n_{\textrm{min}}$, for an
array with $n=14$ spots (note that for the 22-spot solution we do not
reach the $n_{\textrm{min}}$ curve).  In Fig.~\ref{fig444}(b) we
represent the resulting patterns in the space of the concentrations
$(a,b)$.  Again we see that the patterns change continuously as $F$ is
varied, from the blue curve for $F=2.1$ to the red curve for
$F=3.42$.  For the former, a spatial plot of the pattern (inset upper
right) reveals that it is sharply peaked and that a spot has a small
extension. If one perturbs the system by further decreasing $F$ by a
small amount, the number of spots decreases.  Turning our attention to
the other boundary $n_{\textrm{max}}$, an examination of the profiles
there reveals the existence of degenerate values of $a$ for fixed $b$
(marked in red).  This implies that within the spot, a small dip in
the center is formed, as visualized in the inset (lower left).  Now,
as one increases $F$ by a small amount, the spot pattern eventually
splits along this dip.

\section*{Discussion}

In conclusion, using a simple reaction-diffusion model, we have
identified the stability region for $n$-spot solutions in the
parameter space spanned by a natural control parameter (the feed-rate
$F$). In general, for a given $F$, we find multistability of spot
solutions, with a range of spot numbers $n$, bounded by numerically
determined curves $n_{\textrm{min}}$ and $n_{\textrm{max}}$. 

Spot arrays in the reaction-diffusion system~(1) can be created in
very different ways, with two distinct limiting behaviors (single-spot
solution and native Turing pattern). These arrays are
indistinguishable in intermediate regimes (the asymptotic states for
fixed $F$ and $n$ are identical) where both generative mechanisms
coexist.  This means that either mechanism can be used to generate the
same pattern. Therefore, to discriminate between the pattern formation
mechanisms is to some degree artificial, as these can only be
distinguished during their transient phases. However, due to the
different transients in each case, the initial conditions determine
the pattern evolution and the final number of spots in a non-trivial
manner: While small random perturbations create typical Turing
patterns with $n$ coinciding on average with $n_T$, through an
appropriate tuning of $F$, we gain access to a wider range of $n$ via
replication cascades.  As we have shown, one can make use of the
hysteresis feature of the system to generate periodic cycles of spot
replication and destruction.

Despite the simple and specific chemical nature of our model, we
expect the qualitative result to hold for similar non-chemical systems
and in general for those complex scenarios whose dynamics (possibly in
reduced form) can by described by reaction-diffusion models such as
certain fluid
systems~\cite{AvilaS11,ShiPRL13,BarkleyPRE11,MannevilleEpL12}. There,
cycles of spot replication and destruction could be used to engineer
transitions between out-of-equilibrium states. For example, splitting
of turbulent stripes is dominant for large Reynolds numbers in plane
Couette flow, while for low Reynolds numbers stripe decay is
favored~\cite{ShiPRL13}. While the specifics in that system are
different from our model, analogous to the role played by the
feed-rate, we hypothesize that it could be possible to control the
number of stripes through switching of the Reynolds number.

As perspectives for future work we mention the possibility to engineer
the system by modulating the feed-rate in time, using a self-generated
signal (\emph{feedback}) that can use the splitting/disappearance
pathways~\cite{MikhailovPR06}.  Furthermore, transitions between spot
arrays with different $n$ can be also induced by application of
noise. However, the realization of these ideas goes beyond the scope
of this article.

In the spirit of simplicity, tractability and clarity, we have focused
on a medium with one spatial dimension. Obviously, the dynamics of
localized spots and Turing patterns is much richer in two space
dimensions. However, we expect that that the main result of this study
holds qualitatively also for two-dimensional spot arrays.

\section*{Methods}
\label{sec:methods}

The numerical simulations of Eq.~(\ref{eq:diffeqmodel2}) were
conducted in a one-dimensional space of size $L=200$ with periodic
boundary conditions which ensures that there no spots attached to the
boundary (varying $L$ as well as using no-flux boundary conditions
have not shown to produce mayor changes). A spatial grid with $\Delta
x=0.5$ was used along with a Euler routine for time integration and a
3-point stencil for the diffusion operator. In order for increased
accuracy for patterns close to instabilities and for validation
purposes, a 4th-order Runge-Kutta scheme was employed along with a
smaller grid resolution $\Delta x=0.4$ and a 5-point stencil. The
two-dimensional simulations shown in Fig.~\ref{fig:fig111}(a,b) are
only for the purpose of illustration; they correspond to simulations
with $\Delta x=\Delta y=0.5$ and a 5-point stencil for the diffusion
operator.

We are not interested in oscillatory behavior and therefore choose
$k_2=1.2$ and $k_3=1.5$ in order to be far from the Hopf bifurcation
curve (compare Fig.~S1). In order to observe localized spot and Turing
patterns, sufficiently strong substrate diffusion is necessary, and we
set $D_a=1$ and $D_b=50$ accordingly.  Although for one-dimensional
localized patterns, the notation \emph{spike} is used in the
literature, we apply the more general notation \emph{spots}.

To obtain the limiting curves in Fig.~\ref{fig:fig333}, spot solutions
are initialized for different $n$ in the region $F < F_{SN}$.  The
asymptotic state of a simulation is determined at $T=2000$, although
transients usually have died out after $T\approx 10^2$ (if $n$ changes
within the simulation) or $T\approx 10^1$ (if $n$ does not change
within the simulation). Following this, $F$ is increased in increments
of $0.05$, and the simulation is allowed to run again until the
asymptotic state is reached. This procedure is repeated until
splitting is observed. In the same way, $F$ can be either increased
further or decreased until spots split again or disappear. This
iterative process has been exhaustively performed for all possible $n$
to determine the stability area.

We note that the numerical results come with inherent imprecisions, in
particular for large $F$ where the amplitude of the Turing pattern
vanishes and for small $F$ where the spot pattern disappears. Finite
simulation time may mistake a transient for an asymptotic state. Also,
the finite size of the medium (together with the periodic boundary
conditions) implies that the range of $n$ (which is a positive integer
number) is limited. However, simulations for larger system size and
no-flux boundary conditions have not revealed qualitatively new
behavior, though of course $n$ increases and the curves in
Fig.~\ref{fig:fig333} are extensive in system size.

\section*{Acknowledgments}


\section*{Figure Legends}

\end{document}